\documentclass[sigplan,10pt,nonacm]{acmart}
\usepackage{listings}
\usepackage{tabularx}
 
\lstdefinelanguage{lean}{
  morekeywords={example, by, theorem, def, have, intro, rintro, at,
    transfer, transfer_rw, transfer_back, transfer_at, btransfer, Prop},
  sensitive=true,
  morecomment=[l]{--},
}
 
\setcopyright{none}
\acmConference[CPP '27]{Certified Programs and Proofs}{January 11--12,
  2027}{Mexico City, Mexico}

\begin{document}

\title{A Transfer Tactic for Lean}

\author{Zhaoxi Chen}
\email{fourfours1234@gmail.com}
\affiliation{%
  \institution{Independent Researcher}
  \city{Great Neck}
  \country{United States}
}

\author{Daniel Raggi}
\email{dr495@cam.ac.uk}
\affiliation{%
  \institution{University of Cambridge}
  \city{Cambridge}
  \country{United Kingdom}
}

\begin{abstract}
Rewriting in proof assistants spans a strict hierarchy. Equational
rewriting substitutes based on extensional equality;
subequational rewriting substitutes based on homogeneous relations
such as $\leq$ or $\subseteq$, justified by monotonicity
lemmas; and generalized rewriting relates different operations
across different types, justified by transfer rules. All
three are specializations of one schema, the function relator
$(R \Rightarrow S)\,f\,g$. Lean~4's \texttt{rw} and Mathlib's
\texttt{grw}/\texttt{gcongr} implement the first two levels, but the third has
had no Lean implementation. We present one: a \texttt{transfer} tactic
family over a tagged rule database, benchmarked using Mathlib's
own test suites, of which 112 of 178 ported tests close through the
transfer engine. An evaluation of the design's
three motivating hypotheses returns a mixed result:
generality holds, authoring effort wins in its amortized form (averaged
per use), but on dependency footprint, \texttt{transfer} only ties a
mature library and loses on coercions. This shows that
\texttt{transfer}'s value is concentrated on transferring into domains
where few theorems exist.
\end{abstract}



\maketitle

\section{Introduction: three levels of rewriting}
\label{sec:intro}

\subsection{Rewriting with equations}
\label{sec:intro:eq}

Rewriting plays a large role in formal proof. If we know $t_1 = t_2$ and we
know some formula $\phi\,(t_1)$, we should be able to conclude $\phi\,(t_2)$.
Captured as an inference rule,
\[
  \frac{t_1 = t_2\qquad\phi\,(t_1)}{\phi\,(t_2)}\ (\text{rewrite})
\]
is the substitution of equals for equals, sometimes called Leibniz's
law. In practice, the equalities are universally quantified rules such
as $\forall x\, (x + 0 = x)$, and applying one means finding a substitution
z$\sigma$ that maps the rule's left-hand side onto a subterm of the goal.
This process is called pattern matching when only the rule contains
variables, but it is called unification in general.

First order unification is well behaved: when a unifier exists there is a
most general one, unique up to renaming, and it can be computed in
near linear time~\cite{Robinson1965,MartelliMontanari1982}. Higher order
unification, where variables may stand for functions, is not:
second order unification~\cite{Huet1973,Goldfarb1981} and
even third order pattern matching~\cite{Loader2003} are undecidable. Proof
assistants therefore use heuristic procedures that may not terminate,
descending from Huet's semi-decision algorithm~\cite{Huet1975}. The
rule composition machinery of Section~\ref{sec:tool} uses restricted
higher order unification.

\subsection{Rewriting beyond equality}
\label{sec:intro:beyond}

Sometimes we want to substitute under a relation weaker than equality.
Knowing $P \to Q$, we would like to pass from $P \wedge R$ to $Q \wedge R$, or
knowing $A \subseteq B$, we would like to conclude
$A \cup C \subseteq B \cup C$. In each case we ``rewrote'' using relations $\to$ or
$\subseteq$ instead of the stronger $=$. For such an operation to be valid, the surrounding
context must \emph{respect} the relation. We say that a function $f$ respects the relation $S$ provided $S(x,y)$ implies $S(f(x), f(y))$. In our example, the function $(\_ \wedge R)$ respects $\to$ since $P \to Q$ implies $(P \wedge R) \to (Q \wedge R)$. This can easily be generalized to a pair of relations, saying that $f$ respects the pair $R,S$ if $R(x,y)$ implies $S(f(x),f(y))$. The type theoretic account of
rewriting modulo such respectfulness is due to
Sozeau~\cite{Sozeau2009}, whose implementation underlies Coq's
\texttt{setoid\_rewrite}. Moving forward we will use the convention of writing $f\ x$ instead of $f(x)$ and $R\ x\ y$ instead of $R(x,y)$, in line with type-theoretic conventions.

\subsection{The function relator and hierarchy}
\label{sec:intro:hierarchy}

All of the above are instances of one schema. Given relations
$R:\alpha\to\beta\to\mathrm{Prop}$ and
$S:\gamma\to\delta\to\mathrm{Prop}$, the \emph{function relator} $\Rightarrow$ is
defined by
\[
  (R \Rightarrow S)\ f\ g \;:=\;
  \forall x\ y,\ R\ x\ y \to S\ (f\ x)\ (g\ y),
\]
meaning $f$ and $g$ map $R$-related arguments to $S$-related values. (In
Mathlib this is \texttt{Relator.LiftFun}, and multi-argument functions
chain the relator as in $(R \Rightarrow S \Rightarrow T)\,f\,g$.) The
associated elimination rule provides the ability to conclude
$S\,(f\,x)\,(g\,y)$ from $(R \Rightarrow S)\,f\,g$ and $R\,x\,y$:
\[
  \frac{(R \Rightarrow S)\ f\ g \qquad R\ x\ y}{S\ (f\ x)\ (g\ y)}\ (\Rightarrow\!\text{e})
\]

Three specializations of this schema, ordered by the number of parameters
that are allowed to vary, give rise to three levels of rewriting.

\begin{table*}[ht]
\centering
\small
\setlength{\tabcolsep}{5pt}
\begin{tabularx}{\textwidth}{@{}l l >{\raggedright\arraybackslash}X
    >{\raggedright\arraybackslash}X l@{}}
\toprule
Level & Schema instance & Type of fact & Authoring cost & Tactic \\
\midrule
Equational & $({=} \Rightarrow \cdots \Rightarrow {=})\,f\,f$ &
  theorem, for every $f$ & zero & \texttt{rw} \\
Subequational & $(R \Rightarrow R')\,f\,f$ &
  monotonicity, proved and tagged & one lemma per function &
  \texttt{grw} \\
Generalized & $(R \Rightarrow S)\,f\,g$ &
  transfer rule & one rule per pair & \texttt{transfer} \\
\bottomrule
\end{tabularx}
\caption{Three levels of rewriting as three specializations of the function
relator. \texttt{gcongr} is the congruence engine behind \texttt{grw}, and
the \texttt{transfer} family comprises \texttt{transfer},
\texttt{transfer\_rw}, \texttt{transfer\_back} and \texttt{btransfer}.}
\label{tab:hierarchy}
\end{table*}

\paragraph{The equational level} The expression $({=}\Rightarrow\cdots\Rightarrow{=})\,f\,f$ characterizes extensional equality for $f$. It holds for any $n$-ary function $f$, as it simply states that replacing the arguments of $f$ for equals yields equal values. Thus, equational rewriting is a special case of the use of $\Rightarrow$. For instance, the theorem $({=}\Rightarrow{=})\,f\,f$ and the premise $x = y$  can be used to derive $f(x) = f(y)$. If $f(x)$ is a proposition, $f(x) = f(y)$ states equivalence, so $f(x)$ and $x = y$ can be used to conclude $f(y)$. 
In Lean, equational rewriting is implemented in the tactic
\lstinline[language=lean]{rw}.

\paragraph{The subequational level} The expression 
$(R \Rightarrow S)\,f\,f$ 
(often $R=S$) 
states a \emph{monotonicity} fact about $f$ that holds for \emph{some} relations $R,S$. As shown above, this is a theorem when $R$ and $S$ are the equality relation, but otherwise they are not necessarily true. We call this level subequational as it may hold for relations that are weaker than equality such as preorders. In practice, reflexivity closes the untouched argument positions
during congruence descent, while transitivity allows consecutive rewrite
steps to chain. This is the level at which we can use $P \to Q$ and $P \wedge R$ to conclude $Q \wedge R$.
In Lean, subequational rewriting is implemented in the tactics
\lstinline[language=lean]{grw} 
and \lstinline[language=lean]{gcong}.

\paragraph{The generalized level} Expressions of the form $(R \Rightarrow S)\,f\,g$, called \textit{transfer rules}, represent the most general level of rewriting. Because the domains of $f$ and $g$ can be different, the
relations at this level can relate different types, enabling change of
representation transformations. In this paper we present our implementation of a family of tactics that do this.

\subsection{A note on names}
\label{sec:intro:names}

Mathlib's \lstinline[language=lean]{grw} tactic is documented as
``generalized rewriting'' because it generalizes
\lstinline[language=lean]{rw} from $=$ to other relations. In this paper,
however, \lstinline[language=lean]{grw} occupies the subequational level
because it cannot transfer between domains. For the rest of the paper, we
will leave Mathlib's tactic names as they are and use the paper's taxonomy 
when explaining concepts.

\subsection{The Lean 4 landscape}
\label{sec:intro:lean}

Lean~4~\cite{deMouraUllrich2021} and its mathematical library
Mathlib~\cite{mathlib2020} have powerful rewriting tools, but they don't
cover all levels of the hierarchy.

\emph{What exists.} \lstinline[language=lean]{rw} rewrites from equality.
\lstinline[language=lean]{gcongr} (``generalized
congruence'')~\cite{gcongr} applies tagged monotonicity lemmas to reduce a
relational goal to subgoals between the inputs, while side conditions are
handled by an extensible discharger. \lstinline[language=lean]{grw}~\cite{grw}
is the rewriting tool built on \lstinline[language=lean]{gcongr}. Finally,
Lean defines quotient types for equivalence relations.

\emph{What does not exist.} Lean has no user-defined morphisms of the kind
Coq provides via \texttt{setoid\_rewrite}~\cite{Sozeau2009} and no
translation across domains of the kind Isabelle provides via its Transfer
package~\cite{HuffmanKuncar2013}.
\lstinline[language=lean]{rw} implements $({=}\Rightarrow{=})\,f\,f$,
\lstinline[language=lean]{gcongr}/\lstinline[language=lean]{grw} implements
$(R \Rightarrow R')\,f\,f$ for homogeneous relations, and quotients
implement $(Q \Rightarrow Q)$ for quotient surjections $Q$, but there is
currently no implementation of the general $(R \Rightarrow S)\,f\,g$
framework.

\subsection{Contributions}
\label{sec:intro:contributions}

This paper makes three contributions.
\begin{enumerate}
  \item \textbf{An account of the hierarchy}: equational,
        subequational, and generalized rewriting as three specializations
        of one schema.
  \item \textbf{An implementation of the generalized level for Lean~4}
        (Section~\ref{sec:tool}): the
        \lstinline[language=lean]{transfer} tactic family
        (\lstinline[language=lean]{transfer},
        \lstinline[language=lean]{transfer_rw},
        \lstinline[language=lean]{transfer_back}, 
        and \lstinline[language=lean]{btransfer}) over a
        tagged rule database. It consumes the
        \lstinline[language=lean]{@[gcongr]} lemma database directly,
        reuses \lstinline[language=lean]{grw}'s occurrence abstraction,
        and is benchmarked using Mathlib's own test suites.
  \item \textbf{An evaluation} (Section~\ref{sec:eval}) of proof size,
        authoring effort, and generality.
\end{enumerate}
Efforts to automate change of representation transformations have already
been completed in Isabelle~\cite{RaggiEtAl2015,RaggiEtAl2016} using its
Transfer package~\cite{HuffmanKuncar2013}.

\section{The transfer tactic family}
\label{sec:tool}

Our implementation of the generalized level consists of a tagged rule
database and a family of tactics that consume it. The precursor of our
tactic design is Isabelle's Transfer package~\cite{HuffmanKuncar2013}.
Later we will describe how the system relates to Mathlib's
\lstinline[language=lean]{gcongr}/\lstinline[language=lean]{grw}.

\subsection{Transfer rules and attributes}
\label{sec:tool:rules}

A \emph{transfer rule} is a theorem whose statement is a function-relator
fact, registered with the tactic by an attribute. Mathlib's
\texttt{Relator.LiftFun} acts as a foundation for this. The elimination
principle is function application: given the terms
\lstinline[language=lean]{h : (R ⇒ S) f g} and
\lstinline[language=lean]{r : R x y}, the application
\lstinline[language=lean]{h r} proves
\lstinline[language=lean]{S (f x) (g y)}.

Relations that act as bridges between domains, or representations of
mathematical objects, are registered with
\lstinline[language=lean]{@[transfer_relation]}. Standard rules are tagged
with either \lstinline[language=lean]{@[transfer_rule]} or
\lstinline[language=lean]{@[transfer_back_rule]},
and the remaining auxiliary pools are
\lstinline[language=lean]{@[transfer_respects]}, its fallback variant
for decomposing structured predicates, and
\lstinline[language=lean]{@[btransfer_rule]} for the backtracking engine below.
All categories live in a single environment extension, and the rules have
numerical priorities. For example, the cardinality bridge puts its
bijection-reduction rule at priority~2100 so that, to prove equality of
cardinalities, exhibiting a bijection outranks an attempt at direct
equality transfer:
\begin{lstlisting}[language=lean]
@[transfer_back_rule 2100, btransfer_rule]
theorem rel_card_bij (hA : Card A n) (hB : Card B m)
    (h : Nonempty (A ≃ B)) : n = m
\end{lstlisting}
The premise is to reduce an equality of numbers to a bijection
between the witnessing finite sets. The 
\lstinline[language=lean]{transfer_back_rule}
tag serves both the regular and the backtracking engine, but the set priority
only matters to the regular engine, as the backtracking one recovers from the
wrong choice and thus does not need to be steered away from it. 

Within a bridge, rules cluster into recurring shapes: \emph{ground facts}
relating closed forms (\lstinline[language=lean]{BP 1 0}: the number $1$
corresponds to the empty bag); \emph{structural rules} relating operations
argument by argument; \emph{uniqueness and totality facts}
(left/right-unique, left/right-total and their combinations), which
enable equality and existential transfer; and one-way equality
lemmas (\lstinline[language=lean]{rel_eq_mp} needs only
\lstinline[language=lean]{RightUnique R} for the forward direction, and
\lstinline[language=lean]{rel_eq_mpr} only
\lstinline[language=lean]{LeftUnique R} for the backward direction,
refining Mathlib's \lstinline[language=lean]{Relator.rel_eq}).

Crucially, nothing constrains the two function slots to look alike. The
bags-of-primes bridge relates a number to the multiset of its prime
factors:
\begin{lstlisting}[language=lean]
BP : ℕ → Multiset ℕ → Prop
(BP ⇒ BP ⇒ BP) (· * ·) (· + ·) -- mult. ↦ bag union
(BP ⇒ BP ⇒ Iff) (· ∣ ·) (· ≤ ·) -- dvd ↦ multiset order
\end{lstlisting}
This exemplifies the heterogeneity that Section~\ref{sec:intro:hierarchy} identified as
the defining feature of the generalized level.

One family of facts is never tagged: the equational
congruences $({=}\Rightarrow\cdots\Rightarrow{=})\,f\,f$, which hold for
every $f$ at every arity. Tagging them per function would fill the
database with facts that carry no information.

\subsection{A complete bridge}
\label{sec:tool:natint}
 
One bridge in the repository relates a natural number to its
integer coercion, and is small enough to show almost entirely. The
relation and its ground facts form three lines:
\begin{lstlisting}[language=lean]
@[transfer_relation]
def NZ (n : ℕ) (z : ℤ) : Prop := (n : ℤ) = z
 
@[transfer_rule] theorem rel_zero : NZ 0 0 := rfl
@[transfer_rule] theorem rel_one  : NZ 1 1 := rfl
\end{lstlisting}
A structural rule states that an operation respects the bridge, and its
proof is the authoring cost of covering that operation:
\begin{lstlisting}[language=lean]
@[transfer_rule]
theorem rel_add :
    (NZ ⇒ NZ ⇒ NZ) ((· + ·) : ℕ → ℕ → ℕ)
                   ((· + ·) : ℤ → ℤ → ℤ) := by
  intro n₁ z₁ h₁ n₂ z₂ h₂
  change ((n₁ + n₂ : ℕ) : ℤ) = z₁ + z₂
  unfold NZ at h₁ h₂; subst h₁; subst h₂
  exact @Nat.cast_add ℤ _ n₁ n₂
\end{lstlisting}
Not every operation transfers unconditionally. Natural-number subtraction
is truncated ($a - b = 0$ when $a < b$), so the tagged rule carries a precondition that becomes a subgoal at
application time and closes from a matching hypothesis in context:
\begin{lstlisting}[language=lean]
@[transfer_rule]
theorem rel_nat_sub :
    ∀ ⦃n₁ z₁⦄, NZ n₁ z₁ → ∀ ⦃n₂ z₂⦄, NZ n₂ z₂ →
    n₂ ≤ n₁ → NZ (n₁ - n₂) (z₁ - z₂)
\end{lstlisting}
Order transfers through
\lstinline[language=lean]{rel_le : (NZ ⇒ NZ ⇒ Iff) (· ≤ ·) (· ≤ ·)}. The
\lstinline[language=lean]{Iff}-conclusion rule cannot be directly applied to the
goal \lstinline[language=lean]{z₁ ≤ z₂} (the conclusion
\lstinline[language=lean]{_ ↔ _} does not unify with
\lstinline[language=lean]{_ ≤ _}), so the database also includes the arrow
variant \lstinline[language=lean]{rel_le_mp}, derived by
\lstinline[language=lean]{.mp}, which is what backward chaining actually
uses. Finally, four facts fix what we call the bridge's \emph{grade}.
Left-totality, left-uniqueness and right-uniqueness all hold and characterize the bridge
as a partial bijection. Right-totality
fails because \lstinline[language=lean]{NZ} cannot apply to negative integers, which have no preimage.
Which of these theorems are \emph{tagged} is decided not necessarily by their truth, but also by their consequences. \lstinline[language=lean]{RightUnique NZ} unfolds to
\lstinline[language=lean]{b = c} with metavariables on both sides, and
since \lstinline[language=lean]{NZ n z} is definitionally
\lstinline[language=lean]{(n : ℤ) = z}, the rule matches the very
subgoals it generates; tagging it would send the search into a regress. Both
uniqueness directions are therefore kept as proof terms only, and their
conjunction \lstinline[language=lean]{nz_bi_unique : BiUnique NZ} is
tagged in their place, carrying the same content behind a head symbol the
index can discriminate on. In the same way as \lstinline[language=lean]{rel_le_mp}'s constraint, what the
database can hold is shaped by what the search can handle.
 
In use, the rules compose without being named. Nested arithmetic
decomposes through \lstinline[language=lean]{rel_mul} and
\lstinline[language=lean]{rel_add} down to the context hypotheses:
\begin{lstlisting}[language=lean]
example (a b c : ℕ) (a' b' c' : ℤ)
    (ha : NZ a a') (hb : NZ b b') (hc : NZ c c') :
    NZ (a * b + c) (a' * b' + c') := by transfer
\end{lstlisting}
The derivation tree and the tactic call coincide. Below is the same
order-transfer fact, written first as two explicit elimination steps and
then as the tactic that finds them:
\begin{lstlisting}[language=lean]
example (n₁ n₂ : ℕ) (z₁ z₂ : ℤ) (h₁ : NZ n₁ z₁)
    (h₂ : NZ n₂ z₂) (hle : n₁ ≤ n₂) : z₁ ≤ z₂ :=
  (rel_le h₁ h₂).mp hle   -- two (⇒E) steps, then mp
 
example (n₁ n₂ : ℕ) (z₁ z₂ : ℤ) (h₁ : NZ n₁ z₁)
    (h₂ : NZ n₂ z₂) (hle : n₁ ≤ n₂) : z₁ ≤ z₂ := by
  transfer
\end{lstlisting}
\lstinline[language=lean]{transfer_back} performs the
\lstinline[language=lean]{zify}-style translation in reverse: on a
\lstinline[language=lean]{ℕ}-side goal it leaves the \lstinline[language=lean]{ℤ}-side goal, pushing casts through compound
expressions because the structural rules decompose the witness subgoal
\lstinline[language=lean]{NZ (n + m) ?z} into per-argument witnesses
before reflexivity fires. Therefore, \lstinline[language=lean]{?z} unifies to
\lstinline[language=lean]{↑n + ↑m} rather than
\lstinline[language=lean]{↑(n + m)}:
\begin{lstlisting}[language=lean]
example (n m k : ℕ)
    (hz : (↑n : ℤ) + ↑m ≤ ↑k) : n + m ≤ k := by
  transfer_back
  exact hz
\end{lstlisting}

\subsection{The tactics}
\label{sec:tool:tactics}

Four tactics (with variants) consume the database.

\paragraph{\texttt{transfer:}}
Engine-driven backward chaining over the tagged rules and Mathlib's
\lstinline[language=lean]{@[gcongr]} database
(Section~\ref{sec:tool:integration}). The engine uses discrimination
trees, performing indexed candidate lookup with backtracking on failed
rule applications. It tries to
close the goal up to \emph{content residuals}, or subgoals (such as
\lstinline[language=lean]{Nonempty (A ≃ B)}) that are left open. Direct
congruence descent with equality is possible, so ordinary rewriting is
subsumed as the $R = {=}$ instance of transfer. The variant called
\lstinline[language=lean]{transfer to T} skews the search toward a named
target type so that within each priority level, rules whose statement mentions
\lstinline[language=lean]{T} more often are tried first.

\paragraph{\texttt{transfer\_back:}}
Goal translation in the opposite direction. A goal stated in the target
representation is reduced, via the
\lstinline[language=lean]{@[transfer_back_rule]} category, to the
corresponding goal in the source representation, which remains as the
residual (together with any side conditions that the rules could not
discharge). \lstinline[language=lean]{transfer_back to T} is the
corresponding search filter, and \lstinline[language=lean]{transfer_at h}
applies the translation to a hypothesis instead of the goal.

\paragraph{\texttt{transfer\_rw:}}
Generalized rewriting via the transfer engine, with the same idea as
Mathlib's \lstinline[language=lean]{grw}.
\lstinline[language=lean]{transfer_rw [h₁, ← h₂] (at h)?} rewrites with
relational hypotheses \lstinline[language=lean]{hᵢ : Rᵢ aᵢ bᵢ} for any
binary relation, replacing occurrences of $a_i$ with $b_i$ in the goal or
a hypothesis and trying to close by reflexivity. The front end copies the
occurrence abstraction design of
\lstinline[language=lean]{Lean.MVarId.grewrite}, but the implication is
proved by the transfer engine (native rules, the
\lstinline[language=lean]{gcongr} adapter, and a transitivity fallback)
rather than by \lstinline[language=lean]{gcongr} itself. Exactly as in the case of
\lstinline[language=lean]{grw}, Lean's own rewriter is
used for $=$/$\leftrightarrow$ rules. For heterogeneous rules
\lstinline[language=lean]{h : T a b} with
\lstinline[language=lean]{a : α} and \lstinline[language=lean]{b : β} of
different types, where occurrence substitution is poorly defined, the
tactic falls back to goal transfer with \lstinline[language=lean]{h}
seeded as a leaf. Mathlib's \lstinline[language=lean]{grw} is left
available for side-by-side comparison.

\paragraph{\texttt{btransfer:}}
Goal translation with backtracking. Where
\lstinline[language=lean]{transfer} and
\lstinline[language=lean]{transfer_back} accept residuals,
\lstinline[language=lean]{btransfer} requires every subgoal
to close except \lstinline[language=lean]{Nonempty}-headed 
content residuals. This is what makes its backtracking effective:
a wrong rule choice fails downstream instead of succeeding with
a residual. Back-rules come from \lstinline[language=lean]{@[btransfer_rule]}
alongside the shared decomposition pools. 

\subsection{A worked derivation}
\label{sec:tool:worked}

The tactics automate the mechanical composition of elimination steps.
Consider the bags-of-primes bridge with two ground facts and one
structural rule:
\begin{gather*}
  \mathrm{BP}\ 6\ \{2,3\}, \qquad
  \mathrm{BP}\ 120\ \{2,2,2,3,5\},\\
  (\mathrm{BP} \Rightarrow \mathrm{BP} \Rightarrow {\leftarrow})\ (\mid)\ (\leq),
\end{gather*}
where ${\leftarrow}$ stands for backward implication and $\leq$ for multiset order/inclusion.
Applying the elimination rule to the structural rule and the first ground
fact specializes the first argument slot, and applying it again with the
second ground fact specializes the second, leaving
\[
  \{2,3\} \leq \{2,2,2,3,5\} \;\to\; 6 \mid 120 .
\]
The same composition works for any combination of rules. The
\lstinline[language=lean]{transfer} tactics build such derivation trees
automatically. An example on a live goal is that from
\lstinline[language=lean]{Nat.gcd m n ∣ m},
\lstinline[language=lean]{transfer} produces
\lstinline[language=lean]{PF m ∩ PF n ≤ PF m}, where the
gcd-to-intersection and divisibility-to-order steps each come from a
tagged rule.

\subsection{Transfer in a live proof}
\label{sec:tool:live}
 
Quotient bridges are easy to use transfer with. The relation
\lstinline[language=lean]{ZM n : ℤ → ZMod n → Prop} relates an integer to
its residue modulo $n$ and is right-total and right-unique but not
left-unique, meaning equality transfers forward only. With its structural rules tagged, the classic
impossibility argument that no integer squares to 3 runs as a
one-call reduction modulo 5 followed by a finite check:
\begin{lstlisting}[language=lean]
example : ¬ ∃ z : ℤ, z * z = 3 := by
  rintro ⟨z, hz⟩
  have h5 : ((z * z : ℤ) : ZMod 5)
          = ((3 : ℤ) : ZMod 5) := by transfer
  push_cast at h5
  have : ∀ x : ZMod 5, x * x ≠ 3 := by decide
  exact this _ h5
\end{lstlisting}
The \lstinline[language=lean]{transfer} call discharges the casting step, and
the \lstinline[language=lean]{ZM}-multiplication and forward equality
rules compose to carry \lstinline[language=lean]{hz} across the bridge.
What would otherwise be a hand-picked chain of
\lstinline[language=lean]{Int.cast} lemmas is one uniform call, and the
remaining goal is decidable.
 
\subsection{Backward transfer and content residuals}
\label{sec:tool:residuals}
 
In \lstinline[language=lean]{transfer_back}, the engine performs the bookkeeping while the 
mathematical content remains open. The cardinality bridge's
comparison back-rules are an example.
\lstinline[language=lean]{rel_card_bij}
(Section~\ref{sec:tool:rules}) reduces a numeric equality to a bijection
between the witnessing finite sets, and its siblings reduce the
inequalities to an injection or surjection:
\begin{lstlisting}[language=lean]
@[transfer_back_rule, btransfer_rule]
theorem rel_card_inj (hA : Card A n) (hB : Card B m)
    (h : Nonempty (A ↪ B)) : n ≤ m
\end{lstlisting}
On the symmetry of binomial coefficients, $\binom{n}{k} = \binom{n}{n-k}$ for $k \leq n$, the division works
as intended. \lstinline[language=lean]{transfer_back} fires
\lstinline[language=lean]{rel_card_bij} on the equality, decomposes both
sides through the $k$-subset structural rule down to a shared canonical
witness set $S$ of size $n$, and leaves as the sole residual
\begin{lstlisting}[language=lean]
⊢ Nonempty (powersetCard k S ≃
            powersetCard (n - k) S)
\end{lstlisting}
which is discharged by the actual complement bijection, with no
cardinality arithmetic anywhere in the user's proof.
\begin{lstlisting}[language=lean]
def complEquiv [DecidableEq α] (S : Finset α)
    {k m : ℕ} (hS : S.card = m) (hk : k ≤ m) :
    ↥(S.powersetCard k) ≃ ↥(S.powersetCard (m - k))
\end{lstlisting}
The same mechanism unlocks $\binom{n}{k} \leq 2^n$, as the injection rule
reduces it to the inclusion embedding of $k$-subsets into the powerset,
an inequality the forward-only monotonicity route rejects.
 
Pascal's recurrence,
$\binom{n+1}{k+1} = \binom{n}{k} + \binom{n}{k+1}$, marks the
boundary of \lstinline[language=lean]{transfer_back}. 
The combinatorial proof is a
\emph{hybrid} in that the $(k{+}1)$-subsets of \lstinline[language=lean]{insert x s}
split into those not containing $x$ (counted directly) and those
containing $x$ (counted through the \lstinline[language=lean]{insert x}
bijection), so the identity needs the disjoint-union rule and
\lstinline[language=lean]{rel_card_bij} \emph{together}. What
\lstinline[language=lean]{transfer_back} actually does is fire
\lstinline[language=lean]{rel_card_bij} first, after which the
low-priority canonical-witness rule closes the right-hand
\lstinline[language=lean]{Card} witness before the sum can decompose,
and no backtracking revisits this. The residual becomes a bijection
into an arbitrary canonical set of the right size, dischargeable
only by equating cardinalities. The proof state falsely progresses since transfer essentially reduces Pascal to itself. Lifting the
partition lemma to a tagged rule does not change the route; the shortcut
fires before structural decomposition is attempted.
In this case, the tactic's search success criterion is the obstruction.
\lstinline[language=lean]{transfer_back} treats a rule as successful once it
applies and the recursion continues. A choice that leaves a
residual does not fail, so it never triggers a backtrack. On the other hand,
\lstinline[language=lean]{btransfer} 
requires every subgoal to close except 
\lstinline[language=lean]{Nonempty}-headed content residuals, so the wrong
choice fails downstream and the search backtracks. In Pascal's case, it 
backtracks to the partition rule and closes it over the same database.

\subsection{Relationship to \texttt{gcongr} and \texttt{grw}}
\label{sec:tool:integration}

The system began fully independent of Mathlib's subequational level
machinery, containing its own search, rule categories, and rewriting
front end. It is now integrated with Mathlib in three different ways:

\begin{enumerate}
  \item \textbf{Database}: The engine consumes Mathlib's
        \lstinline[language=lean]{@[gcongr]} lemma database in the
        following way: an adapter queries the
        \lstinline[language=lean]{gcongr} environment extension with its
        own key, making roughly 700 monotonicity lemmas usable
        without re-tagging. The \lstinline[language=lean]{gcongr} tactic
        itself is never called on goals, but its discharger is used for
        side conditions such as positivity.
  \item \textbf{Skeleton}:
        \lstinline[language=lean]{transfer_rw} reuses
        \lstinline[language=lean]{grw}'s occurrence abstraction, with the
        transfer engine proving the resulting implication. The earlier
        rewriting front end was retired in its favor.
  \item \textbf{Benchmark}: The system is passed through Mathlib's own
        test suites for \lstinline[language=lean]{gcongr} and
        \lstinline[language=lean]{grw}, ported to a parity suite
        (Section~\ref{sec:eval:h3}). In the deferred cases, parity files
        still run the original tactics, so both systems remain live.
\end{enumerate}

A summary is that the transfer engine subsumes
\lstinline[language=lean]{grw}/\lstinline[language=lean]{gcongr} by
consuming its rule database and reproducing its test suites. It is neither
a parallel system nor a wrapper around Mathlib's tactics. This framing
clears up the generality comparison of Section~\ref{sec:eval}, which
uses the same lemma database and the same tests.

\subsection{Outlook: refinement is a bridge}
\label{sec:tool:refinement}
 
A data refinement relation is a bridge. Where the transformations of this
paper relate two representations of a mathematical object, a refinement
relation \lstinline[language=lean]{R : Impl → Spec → Prop} relates a
program state to the abstract value it implements; operations preserved
by the refinement are exactly relator facts
$(R \Rightarrow R)\,\mathit{op}_{\mathrm{impl}}\,\mathit{op}_{\mathrm{spec}}$,
observations are $(R \Rightarrow {=})$ facts, and invariant-guarded
operations take the precondition shape of
\lstinline[language=lean]{rel_nat_sub}
(Section~\ref{sec:tool:natint}). The classic batched queue is a pair of
lists refining a single list, with amortized-constant enqueue. We sketch the bridge in the framework's notation, with each rule
being a two-line list lemma:
\begin{lstlisting}[language=lean]
@[transfer_relation]
def RQ (p : List α × List α) (q : List α) : Prop :=
  p.1 ++ p.2.reverse = q
 
@[transfer_rule] theorem rel_empty : RQ ([], []) []
 
@[transfer_rule]  -- enqueue: cons on the back list
theorem rel_enq (x : α) :
    (RQ ⇒ RQ) (fun p => (p.1, x :: p.2))
              (fun q => q ++ [x])
 
@[transfer_rule]  -- dequeue, guarded by nonemptiness
theorem rel_deq :
    ∀ ⦃p q⦄, RQ p q → p.1 ≠ [] →
    RQ (p.1.tail, p.2) q.tail
 
theorem rebalance :  -- two states, one queue
    ∀ ⦃b q⦄, RQ ([], b) q → RQ (b.reverse, []) q
\end{lstlisting}
The relation's grade is interesting. \lstinline[language=lean]{RQ}
is left-total (every pair abstracts to something), right-total (every
queue is implemented, by \lstinline[language=lean]{(q, [])}), and
right-unique (each state implements one queue), but not left-unique, as
\lstinline[language=lean]{rebalance} exhibits two states implementing the
same queue. That is the grade of the quotient row of
Table~\ref{tab:hierarchy}, and the consequences are that
equality transfers from implementation to specification only, and two
states with equal abstractions are observationally equal without actually
being equal. Refinement is the quotient pattern read operationally.
Proofs then flow in the refinement direction.
\lstinline[language=lean]{transfer_back}, on an implementation-side goal,
leaves the specification-side goal, so correctness is proved against
the abstract queue once and then transported to the batched one.
 
The bridge above is a sketch
and is not part of the corpus evaluated in
Section~\ref{sec:eval}. The identification itself is established
practice elsewhere---Isabelle's data refinement is built on the same
Lifting/Transfer machinery~\cite{HuffmanKuncar2013,HaftmannEtAl2013},
and CoqEAL's ``refinements for free'' brings
refinement-by-relations to Coq with the same relator
discipline~\cite{CoqEAL2013}. The point of the sketch is that the
framework of this paper expresses it with no new machinery, and a refinement
is one more entry in the bridge catalogue.

\section{Evaluation}
\label{sec:eval}

The design was motivated by three hypotheses:
\begin{description}
  \item[H1 (proof size).] Transfer shortens proofs that cross domains.
  \item[H2 (authoring effort).] Transfer reduces the investment needed to
        build proofs.
  \item[H3 (generality).] Transfer subsumes
        \lstinline[language=lean]{rw}/\lstinline[language=lean]{grw}/%
        \lstinline[language=lean]{gcongr}.
\end{description}
We evaluate each against the repository's corpora: a baseline of nine
bridge modules with 118 in-file examples, single-theorem case studies,
two break-even families measuring dependency footprint as a corpus grows,
a pre-registered grid on the coercion class, and a 178-test parity suite
ported from Mathlib's own \lstinline[language=lean]{gcongr} and
\lstinline[language=lean]{grw} tests. For each hypothesis we draw a
verdict after presenting the evidence for and against.

\paragraph{Methodology.}
We report three metrics that have different purposes. \emph{Raw proof
length}, which counts the lines of tactic script or proof-term nodes, is
the visible cost at the use site. \emph{Distinct-constant count} is the
number of distinct constants a proof term references. \emph{Proof-graph
size}, written $|G|$, is the transitive closure of constants used by a
proof, terminating at axioms. The third metric measures true dependency
footprint and is the only one on which we judge H1, but even there,
the granularity matters. A single theorem's proof-graph size is dominated
by the route its proof takes rather than by the tactic that found it,
so only cumulative comparisons that use the same route carry evidential
weight.
However, raw length is completely
dominated by strategy and can hide true cost in both directions, as a
one-line \lstinline[language=lean]{by transfer} conceals the bridge's
fixed rule set while a short hand proof can conceal a detour through the
typeclass instance tower. The first two metrics remain informative about
\emph{authoring} effort (H2), where the use site cost is at issue, with
the caveat just stated for the first. For footprint we separate
\emph{fixed} cost (the bridge) from \emph{marginal} cost (each new
theorem), and the headline metric is amortized---a bridge of $R$ rules closing
$T$ theorems costs $R/T$ rules per theorem, and we track whether that
investment pays off as cumulative union proof-graph size $|G|$ against
corpus size $k$.

\subsection{H1: proof size and dependency footprint}
\label{sec:eval:h1}

\paragraph{Evidence for.}
Three cardinality theorems proved by \lstinline[language=lean]{transfer}
have individual proof graphs of 4483, 2565, and 3431 declarations, but due to reuse,
the size of their combined graph is only 4739. 
Theorems using the same bridge
share a big part of their whole graph, so the marginal cost of an
additional theorem collapses. 

\paragraph{Evidence against.}
The break-even experiments return negatives. Six distinct cardinality
identities, each proved on the same goal by both
\lstinline[language=lean]{transfer} and a hand proof using Mathlib's
\lstinline[language=lean]{card_*} lemmas, give cumulative union
proof-graphs that are nearly identical. The six identities count
powersets, products, and their compositions:
$|\mathcal{P}(s)| = 2^n$, $|s \times t| = nm$,
$|\mathcal{P}(s \times t)| = 2^{nm}$,
$|\mathcal{P}(s) \times \mathcal{P}(t)| = 2^n \cdot 2^m$,
$|\mathcal{P}(\mathcal{P}(s))| = 2^{2^n}$, and
$|\mathcal{P}(s) \times t| = 2^n \cdot m$.

\begin{table*}[ht]
\centering
\small
\begin{tabular}{@{}rrr@{\qquad}|@{\qquad}rrr@{}}
\toprule
\multicolumn{3}{c}{\texttt{Card} family} & \multicolumn{3}{c}{\texttt{BP} family} \\
$k$ & transfer $|G|$ & incumbent $|G|$ & $k$ & transfer $|G|$ & incumbent $|G|$ \\
\midrule
1 & 2565 & 2564 & 1 & 418 & 417 \\
2 & 2905 & 2902 & 2 & 419 & 418 \\
3 & 2906 & 2903 & 3 & 420 & 419 \\
4 & 2907 & 2904 & 4 & 431 & 429 \\
5 & 2908 & 2905 & 5 & 432 & 430 \\
6 & 2909 & 2906 & 6 & 433 & 431 \\
\bottomrule
\end{tabular}
\caption{Break-even experiments: cumulative union proof-graph size as the
corpus grows, transfer vs.\ hand proofs from library lemmas
(``incumbent''---the established route the library already provides). No
crossover on either family.}
\label{tab:breakeven}
\end{table*}

The same experiment on the structurally different bags-of-primes bridge
(non-injective, predicated totality, multiset-valued) reproduces the
result. A bridge's rules are proved from the same library lemmas an
incumbent hand proof uses, so the two dependency graphs share their core.
A footprint win therefore requires a domain where the incumbent reinvents
the machinery of each theorem---which does not arise when the incumbent
has reusable lemmas, as is the common case in Mathlib.

The coercion class is another negative. On a pre-registered $10$-row grid
over relations $\{\leq, =\}$ and expression shapes (atom, sums, products,
nesting), \lstinline[language=lean]{transfer_back} closes all ten rows,
but so do \lstinline[language=lean]{exact_mod_cast} and
\lstinline[language=lean]{omega}, identically. Meanwhile, transfer
requires the $\mathbb{N}\to\mathbb{Z}$ bridge (531 lines, 16 rules) as a
fixed cost that the built-ins do not. On bi-unique coercions, where
\lstinline[language=lean]{norm_cast} is complete, the bridge is purely
overhead. Finally, a decision procedure dominates on decidable arithmetic
fragments, as \lstinline[language=lean]{omega} proves a representative
cast goal with a 133-node term against transfer's 33497.

Two audits qualify the single-theorem comparisons in both directions. The
binomial gap decomposes into a shared core of 3245 declarations plus 1238
unique to transfer (but only about 9 from the transfer framework itself) and
396 unique to the incumbent. However, the incumbent uses an algebraic
shortcut through the ring hierarchy, an example of how the tactic can be
conflated with the mathematical strategy. Additionally, proof graph
footprint is highly sensitive to proof style, as a single typeclass lemma
(\lstinline[language=lean]{add_zero} resolving through an instance tower)
added over 1400 declarations into a hand proof before being swapped for
the targeted multiset lemma. Fair comparisons must hold the style fixed,
which is something only satisfied by Table~\ref{tab:breakeven}.

\paragraph{Why single-theorem footprint comparisons do not settle H1}
The proof-graph size of $2^n = \sum_k \binom{n}{k}$ via 
\lstinline[language=lean]{transfer_back} is 4483 compared to 3641 for 
Mathlib's incumbent. However, in raw proof length,
\lstinline[language=lean]{transfer_back} wins
by a factor of five. It uses 751 proof-term nodes referencing 36 distinct constants,
compared to 3718 nodes and 58 constants for Mathlib's algebraic
\lstinline[language=lean]{Nat.sum_range_choose}. Neither the proof-graph
footprint nor the size at the use site is evidential for the tactic. The two graphs
share a core of 3245 declarations, with 1238 unique to transfer and 396 unique
to the incumbent.  The remainder is the 
\lstinline[language=lean]{List}/\lstinline[language=lean]{Multiset}
substrate that Transfer's combinatorial route requires, set against the
ring hierarchy required by the incumbent's algebraic proof through 
\lstinline[language=lean]{add_pow}. One generic monoid lemma (\lstinline[language=lean]{add_zero}, 
resolving through \lstinline[language=lean]{Multiset}'s instance tower)
adds 1437 declarations, showing that footprint is just as sensitive
to style. Only same-style cumulative comparisons
are meaningful in this respect, which is what Table~\ref{tab:breakeven} 
measures and what we judge H1 on.

\paragraph{Verdict.}
H1 fails as stated on well-developed targets: transfer ties the incumbent
on footprint and loses on bi-unique coercions. Where the target theorem
already exists, transfer's advantage is authoring, not footprint (H2),
and where it does not exist, there is no incumbent to amortize against at
all. Transfer is better judged as a transport tool than as a refactoring
tool.

\subsection{H2: authoring effort}
\label{sec:eval:h2}

\paragraph{Evidence for.}
Across nine bridges, the tagged rule categories total 114 forward rules
(plus 6 low-priority rules and 26 backward rules). The bridges themselves
total 1527 lines of code (3223 raw, including documentation and in-file
tests) for 117 tagged transfer lemmas. This investment serves the
118-example baseline and the entire parity suite, and the amortization
results show $T$ grows at near-zero marginal footprint, so $R/T \to 0$
for a well-exercised bridge. With the \lstinline[language=lean]{gcongr}
adapter, about 700 of Mathlib's existing monotonicity lemmas become
usable without re-tagging. Authoring style also entails the advantage that
bridges supply targeted ground facts
(\lstinline[language=lean]{rel_one : BP 1 0}), which structurally avoids
the instance-tower detours that inflated the previous hand proof by 1400
declarations. Finally, the use site is often one uniform line with no
need to know which lemma chain applies.

\paragraph{Evidence against.}
Writing correct rules requires understanding the relator shapes, and even
then it often needs iteration. Priorities are also difficult to set. For
example, the cardinality bridge's bijection rule must be tagged at a
priority that outranks the generic equality route, but this can only be
known by seeing the search fail. This can be done only by understanding
the computational tools used, not the mathematical semantics. This
perhaps suggests the need to weight rules in some systematic, automatic
way, but at the moment this is infeasible.

\paragraph{Verdict.}
H2 holds in its amortized form, with an expertise barrier in rule
authoring. Similar to H1, the benefit is greatest where no reusable
incumbent library exists.

\subsection{H3: generality}
\label{sec:eval:h3}

\paragraph{Evidence for.}
In principle, the hierarchy of Section~\ref{sec:intro:hierarchy} settles
the containment: the $f = g$ constraint confines
\lstinline[language=lean]{grw}/\lstinline[language=lean]{gcongr} to
homogeneous relations, and every heterogeneous bridge in the repository
is outside their reach by construction. For a hypothesis
\lstinline[language=lean]{h : T a b} across different types, occurrence
substitution is poorly defined and \lstinline[language=lean]{grw} cannot
be applied, while \lstinline[language=lean]{transfer_rw} falls back to
goal transfer. On the equational side, ordinary
\lstinline[language=lean]{rw}'s proof term is exactly the congruence
$({=}\Rightarrow\cdots\Rightarrow{=})\,f\,f$ applied to the rewrite
equation, so rewriting is the $R = {=}$ instance.

In practice, the claim is tested by porting Mathlib's own test suites for
\lstinline[language=lean]{gcongr} and \lstinline[language=lean]{grw}, with
\lstinline[language=lean]{gcongr} replaced with \lstinline[language=lean]{transfer}
and \lstinline[language=lean]{grw} with
\lstinline[language=lean]{transfer_rw}.

\begin{table*}[ht]
\centering
\small
\begin{tabular}{@{}lrrrrr@{}}
\toprule
File & Tests & Via transfer & Def.\ (UX) & Def.\ (partial) & Fail \\
\midrule
GCongrBasic        & 10 &  7 &  3 &  0 & 0 \\
GCongrCore         & 12 &  5 &  7 &  0 & 0 \\
GCongrImplications & 12 &  5 &  7 &  0 & 0 \\
GCongrInequalities & 74 & 42 & 20 & 12 & 0 \\
GCongrMod          & 17 & 11 &  6 &  0 & 0 \\
GRewrite           & 53 & 42 & 11 &  0 & 0 \\
\midrule
\textbf{Total} & \textbf{178} & \textbf{112} & \textbf{54} & \textbf{12}
  & \textbf{0} \\
\bottomrule
\end{tabular}
\caption{Parity suite: Mathlib's \texttt{gcongr} and \texttt{grw} test
suites replayed through the transfer engine. Each test either passes via
transfer or carries an explicit deferral tag with the original tactic
retained; there are no engine failures.}
\label{tab:parity}
\end{table*}

Of 178 ported tests, 112 close through the production engine and none
fail outright, meaning each deferral is a gap in the interface rather than in
the engine. The full passes include mixed strict/non-strict rewrite
chains, $\mathbb{Q}\to\mathbb{R}$ casts, divisibility rewrites, the
\lstinline[language=lean]{at *} wildcard with correct self-rewrite
skipping, and rewriting through a sum-subset chain. In five sibling tests
the port improved on the original. Side goals that
\lstinline[language=lean]{gcongr} left to \lstinline[language=lean]{simp}
closed under the engine's own discharger, collapsing the proofs to a bare
transfer.

The following is a representative ported test from the transparency section of Mathlib's
\lstinline[language=lean]{gcongr} suite. The test's own
\lstinline[language=lean]{@[gcongr]}-tagged lemma is consumed through the
adapter, and the finishing call---originally by
\lstinline[language=lean]{gcongr}---becomes
\lstinline[language=lean]{by transfer} with no other change.
\begin{lstlisting}[language=lean]
@[gcongr] lemma mono_measure (h : ν ≤ ν') :
    ν a ≤ ν' a := h a
 
example (h : ν ≤ ν') : ν a ≤ ν' a := by
  transfer   -- original: by gcongr
\end{lstlisting}

\paragraph{Evidence against.}
The 54 UX deferrals mark where \lstinline[language=lean]{gcongr}'s
interface is still ahead: template patterns with
\lstinline[language=lean]{?_} holes for depth control (22 tests),
asserted error message text (10), the \lstinline[language=lean]{rel}
finishing variant (7), and smaller families for implication-hypothesis,
binder-naming intro patterns, explicit recursion depth, occurrence
configuration, and $n$th-occurrence rewriting. The 12 partial progress
deferrals expose the consequences of the transfer tactics having an
all-or-nothing principle. Finally, a search based tactic has unpredictable failure
modes, requiring understanding of the rule database for diagnosis.

\paragraph{Verdict.}
H3 holds at the engine level. Subsumption is demonstrated on Mathlib's
own tests over the same lemma database, with no engine failures,  and
heterogeneity goes further beyond the subequational tactics.

\subsection{Summary}
\label{sec:eval:summary}

Generality (H3) is a win because the engine reproduces the subequational test
suites and expresses the heterogeneity that other tactics lack. Authoring
effort (H2) wins in its amortized form, bounded by an expertise cost in
rule authoring. Proof size (H1) is a tie on well-developed targets and a
loss on bi-unique coercions, and its failure shows that transfer's value
concentrates on transferring into domains where few or no target theorems
already exist, not on re-deriving what a mature library already proves well.

\section{Related work}
\label{sec:related}

\paragraph{Transfer in Isabelle/HOL.}
The design of the tactics presented in this paper was inspired by 
Isabelle's Transfer package~\cite{HuffmanKuncar2013}, which introduced 
the tagged-relator paradigm implemented in this paper for Lean, and also from 
its use for data refinement~\cite{HaftmannEtAl2013}. 
The distinction comes from the deep differences between Isabelle and 
Lean. Isabelle's package serves as a foundation for quotient types in
HOL, whereas Lean has native \lstinline[language=lean]{Quot} for quotient 
types. The question is how an implementation of transfer in Lean 
earns back its cost given the preexisting \lstinline[language=lean]{Quot} machinery.

\paragraph{Parametricity-based transfer in Coq.}
A different approach computes the transfer from the statement's type structure 
rather than from tagged facts. Univalent parametricity transforms based on type
equivalences~\cite{TabareauEtAl2018,TabareauEtAl2021}, and
Trocq~\cite{Trocq2024} generalizes it to a hierarchy of relation classes,
so a transfer requires only as much structure as the statement demands
and univalence is avoided where it is not needed. CoqEAL applies the same
relator discipline to program refinement~\cite{CoqEAL2013}. 
A translation is automatic and complete over the fragment it covers, but
constrains which relations may enter it. A search over tagged facts
accepts any heterogeneous relation, including bridges whose grade is
neither an equivalence nor a function, but finds only what has been
tagged.
Trocq's classification of relations by how much
equivalence structure they carry is, in spirit, close to the totality and
uniqueness grade used in this paper.

\paragraph{Cast handling in Lean.}
\lstinline[language=lean]{norm_cast} and its family normalize coercions
through a dedicated simp-set discipline~\cite{LewisMadelaine2020}. On
bi-unique coercions, this is the incumbent that transfer does not beat (the
coercion grid closes under \lstinline[language=lean]{exact_mod_cast} as
completely as under \lstinline[language=lean]{transfer_back}, and the
bridge is pure overhead).

\paragraph{Syntactic transport in Mathlib.}
\lstinline[language=lean]{to_additive}~\cite{mathlib2020} is one of Mathlib's
own transfer mechanisms that is its most heavily used. It is
complementary rather than competing: the correspondence is fixed and
built in, the translation is syntactic and runs once at declaration time,
and it produces new declarations rather than closing goals. The mechanism
here is semantic, user-extensible, and applied to a goal mid-proof.

\section{Conclusion}
\label{sec:conclusion}

\subsection{Summary}

This paper set out a strict hierarchy of rewriting---equational,
subequational, generalized---as three specializations of one schema, the
function relator $(R \Rightarrow S)\,f\,g$, ordered by the number of
parameters that are allowed to vary and, in consequence, by what each
level entails in authoring cost: nothing, one lemma per function, one
rule per pair of operations; in that order. Lean's
\lstinline[language=lean]{rw} and Mathlib's
\lstinline[language=lean]{grw}/\lstinline[language=lean]{gcongr} occupy
the first two levels, and the \lstinline[language=lean]{transfer} family
presented here implements the third, the first such implementation for
Lean~4. It is not a parallel system, as the engine consumes Mathlib's
\lstinline[language=lean]{@[gcongr]} database directly, reuses
\lstinline[language=lean]{grw}'s occurrence abstraction, and takes
Mathlib's own test suites as its acceptance bar.

The evaluation returned one clean win, one qualified win, and one
instructive failure. Generality (H3) holds at the engine level: 112 of
178 ported Mathlib tests close through the transfer engine, and
heterogeneity (different operations across different types) is beyond the
subequational tactics in principle, not merely in practice. Having many (a hundred) tagged rules could in turn
serve hundreds of goals, so authoring
effort (H2) holds in its amortized form but with an expertise barrier at rule-writing
time. Proof size (H1), measured by dependency footprint, is a tie against
a mature library and a loss on bi-unique coercions, where
\lstinline[language=lean]{norm_cast} is already complete. A bridge's
rules are proved from the same lemmas a hand proof uses, so footprint
parity is structural, and transfer's value concentrates where the
incumbent must reinvent each theorem.

\subsection{Future work}

Three directions follow directly from the evaluation.

\emph{Interface parity.} Each parity deferral is an interface gap with an
identified path: template patterns for depth control, the
\lstinline[language=lean]{rel} finishing variant, occurrence
configuration, and---most substantively---a progress-tactic mode. The
engine already carries the required contract internally (main versus side
subgoals, loose side-constraints) for
\lstinline[language=lean]{transfer_rw}; exposing it as a
\lstinline[language=lean]{transfer}-variant syntax would convert the
twelve partial-progress deferrals.

\emph{Transport targets.} The reframing that closed
Section~\ref{sec:eval:h1} implies that the next thing to do is to
evaluate transfer where the target theorem does not yet exist---transport
into under-developed or better-behaved domains---since there the
break-even question against an incumbent does not arise. The boundary
drawn by the footprint studies is also a map of where to look.

\emph{Further developments.} Subsequent work on the same engine, outside
the scope of this report, explores treating a transfer relation together
with its tagged rules as an executable \emph{analogy} between domains:
inspecting the correspondence a bridge induces, synthesizing the analogue
of a given theorem rather than proving a stated one, sweeping a namespace
for its transferable fragment, and composing bridges.

\subsection{Big picture}

The ingredients for the generalized level were already present in Lean. There was a
typeclass-style attribute system to play the role of the rule database,
an existing corpus of some 700 tagged monotonicity lemmas, and a
metaprogramming interface expressive enough to build derivation trees.
What was missing was the unifying schema. Seen through it, ordinary
rewriting, congruence of inequalities, quotient lifting, and cross-domain
transfer are not four mechanisms but one, at different settings of the
same three parameters---and now a library that has already established its
subequational level rule database gets the generalized level's engine
over the same rules.

\begin{acks}
The implementation, the evaluation corpora, and drafts of this paper were
developed with substantial assistance from Claude (Anthropic), used as a
pair-programming and drafting tool under the authors' direction, and the
authors take full responsibility for the content.
\end{acks}

\bibliographystyle{ACM-Reference-Format}
\bibliography{references}

\end{document}